\documentclass[preprint,5p,times,twocolumn]{elsarticle}

\usepackage[utf8]{inputenc}
\usepackage{graphicx}
\usepackage{siunitx}
\usepackage{subcaption}
\usepackage{booktabs}
\usepackage{tabularray}
\usepackage{rotating}
\usepackage{multirow}
\usepackage{natbib}
\usepackage{amsmath}
\usepackage[hidelinks]{hyperref}

\newcommand\orcid[1]{\href{https://orcid.org/#1}{\textsuperscript{ID}}}

\journal{Geoderma}

\begin{document}

\begin{frontmatter}

\title{Predicting Neutron Attenuation from Bulk Density and Moisture for Soil Carbon Measurement}

\author[inst1]{William Larsen\orcid{0000-0003-2001-9860}\fnref{fn1}}
\affiliation[inst1]{organization={Accelerator Technology and Applied Physics Division, Lawrence Berkeley National Laboratory}, addressline={1 Cyclotron Road}, city={Berkeley}, postcode={94720}, state={CA}, country={USA}}

\author[inst1]{Valerie Smykalov\orcid{0000-0001-6290-197X}\fnref{fn2}}

\author[inst2]{Cristina Castanha\orcid{0000-0001-7327-5169}}
\affiliation[inst2]{organization={Earth and Environmental Sciences, Lawrence Berkeley National Laboratory}, addressline={1 Cyclotron Road}, city={Berkeley}, postcode={94720}, state={CA}, country={USA}}

\author[inst2]{Eoin Brodie\orcid{0000-0002-8453-8435}}

\author[inst1]{Mauricio Ayllon Unzueta\orcid{0000-0001-7109-169X}}

\author[inst1]{Bernhard Ludewigt\orcid{0000-0001-9475-0678}}

\author[inst1]{Arun Persaud\orcid{0000-0003-3186-8358}\corref{cor1}}
\cortext[cor1]{Corresponding author: apersaud@lbl.gov}

\fntext[fn2]{Now at: Terradot, USA.}
\fntext[fn2]{Now at: Department of Civil and Environmental Engineering, Penn State University, University Park, PA, USA.}

\begin{abstract}
Inelastic neutron scattering (INS) enables rapid, non-destructive \textit{in situ} measurements of soil elemental composition over large soil volumes. Standard INS yields bulk elemental concentrations, but spatially resolved measurements require techniques such as Associated Particle Imaging (API), which pairs neutron detection with coincident alpha detection to reconstruct the location of the neutron interaction. 
One of the unique advantages of API is its capability to measure all major soil components simultaneously, allowing for the estimation of both bulk density and water content directly from the measured neutron-induced gamma-ray spectra.
Accurate interpretation of bulk INS–API data depends on correcting for both gamma-ray and neutron attenuation in soil. Although gamma attenuation can be calculated from known mass attenuation coefficient data and density, neutron attenuation is more complex, depending on neutron energy, soil composition, bulk density, and hydrogen content from water and organic matter. We use Monte Carlo simulations of soils with varied compositions, bulk densities, and water contents to model neutron attenuation and develop a simple predictive model requiring only dry bulk density and volumetric water content. We validate this model experimentally using an INS–API system with controlled soil columns, finding agreement within 10\% at \qty{30}{\cm} depth.
This approach enables practical, field-ready correction of INS–API measurements for neutron attenuation, laying the groundwork for a self-consistent measurement framework that can address the elemental composition of soil carbon assessments.
\end{abstract}


\end{frontmatter}

\section{Introduction}
Soil carbon (C) is a key component of soil health \cite{lal2014societal} and one of the largest terrestrial reservoirs of C, exchanging rapidly with atmospheric CO$_2$ \cite{padarian2022soil}. Current methods for determining soil C rely on point sampling of soil cores and laboratory analysis, which are time-consuming, expensive, and spatially limited.  

Prompt gamma rays offer a non-destructive, field-based alternative for measuring elemental composition in soils \cite{Wielopolski2011-nq,ayllon2021all,unzueta2022repeatable}. High-energy (\qty{14.1}{\MeV}) neutrons are directed toward the soil, where they collide with nuclei, exciting them and producing characteristic prompt gamma rays upon de-excitation. Inelastic Neutron Scattering (INS) is the dominant mechanism for the production of prompt gamma rays, although other nuclear processes can also contribute to prompt signals. For carbon, the dominant signature is a \qty{4.43}{\MeV} gamma ray \cite{wielopolski2000soil}. By summing counts from such characteristic gamma rays over the irradiated volume, elemental concentrations can be calculated, given appropriate calibration.  

INS integrates over large soil volumes (\qty{\sim0.3}{\m\cubed}) \cite{wielopolski2008nondestructive}, typically sensing to \qty{\sim 30}{\cm} depth \cite{wielopolski2012nuclear,Yakubova2022-bq,Yakubova2025measuring}, but with greatest sensitivity in the upper \qtyrange{0}{10}{\cm} \cite{yakubova2016benchmarking}. This depth dependence reflects attenuation of both high-energy neutrons and the emitted gamma rays.  

\begin{figure}[ht]
\centering
\includegraphics[width=0.8\linewidth]{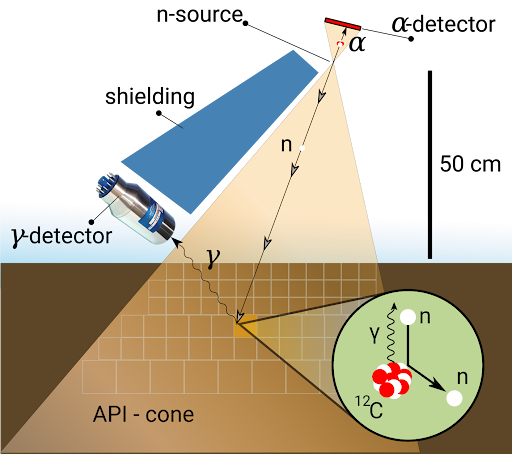}
\caption{Schematic of the INS–API system showing the emission of a \qty{14.1}{\MeV} neutron and associated \qty{3.5}{\MeV} alpha particle from the D–T reaction. The alpha particle is detected by a YAP scintillator, providing the neutron’s initial trajectory and time of emission, while gamma rays from neutron interactions in the soil are detected by LaBr$_3$ and NaI scintillators.}
\label{fig:schematic}
\end{figure}

Associated Particle Imaging (API) enhances INS by detecting the alpha particle emitted in coincidence with each neutron from the D–T fusion reaction \cite{ayllon2021all}. The alpha provides the neutron’s initial trajectory, and, combined with the gamma detection time, allows reconstruction of the scattering location in three dimensions with centimeter resolution. This enables voxelated 3D imaging of elemental distributions to \qty{\sim 30}{\cm} depth. In contrast to combustion or spectroscopic techniques that rely on point sampling of small volumes \cite{Sorenson2018-cm}, INS-API allows for completely non-destructive sampling of large samples (\qty{\sim100}{kg} of soil). A schematic of an API-INS system is shown in Fig.~\ref{fig:schematic} showing the main components of the system, the voxelated volume in the soil. The position of the alpha detector and the small point-like source of DT-fusion reactions define a volume, the API-cone, in the soil/target area that the system is sensitive to.

Recent advances in soil carbon mapping algorithms can achieve a much finer spatial resolution over large regions \cite{De_Gruijter2016-hf}, but their accuracy depends on dense, high-quality ground-truth measurements \cite{Viscarra_Rossel2016-rv, Stanley2023-qr}. Current core sampling and laboratory analysis workflows do not scale well to the large number of measurements needed within smaller areas, limiting the reliability of such models while also requiring extra measurements such as bulk density. Fast, non-destructive methods such as INS–API are therefore critical for providing the spatially rich datasets these models require.

Accurate bulk INS–API quantification, however, requires correcting for gamma-ray attenuation (readily calculated from density and mass attenuation coefficients data) and neutron attenuation, which depends strongly on hydrogen content from water and organic matter \cite{wielopolski2005basic}. Without such a correction, the prediction of the C content for deeper voxels will be inaccurate.  This is especially important when higher resolution voxelation is needed. Kavetskiy et al.~\cite{Kavetskiy2024-ou} showed that using only the timing signal from the alpha particles (i.e., no $XY$ voxelation and only a few depth layers) already requires attenuation correction. Their approach uses corrections to the fitting algorithm of  measured gamma spectra based on simulations of a limited range of soil and water contents, whereas in this work we explore a wider variety of soils and range of water contents to build a more general model based on water/hydrogen contents from different sources in soil.

Monte Carlo neutron transport simulations, such as the Monte Carlo N-Particle (MCNP) software, \cite{goorley2012initial} can model particle attenuation processes, incorporating the effects of soil composition, total atomic density, and hydrogen content. In this work, we simulate neutron attenuation in soils with varied bulk densities, water contents, and chemical compositions, including hydrogen from free water, lattice water, and organic matter. We fit an empirical model relating the attenuation of INS reaction rates to only bulk density and water content and validate it experimentally using an INS–API system with controlled soil columns \cite{ayllon2021all,unzueta2019associated}.

In upcoming publications, we aim to extend our findings by demonstrating how the estimation of bulk density and moisture content from INS–API measurements leads to a self-consistent model for elemental composition across varied soil types. However, this paper primarily focuses on the development and validation of neutron attenuation models. By establishing a robust framework for attenuation correction, we lay the groundwork for integrating bulk density and moisture estimations into a comprehensive approach for enhancing soil elemental assessments.

\section{Materials and Methods}

\subsection{Gamma-ray attenuation coefficients}
The attenuation of gamma rays through soil is described by the mass attenuation coefficient ($\mu_{m}$, \si{\centi\meter\squared\per\gram}), which quantifies the probability of gamma interaction per unit mass. We retrieved $\mu_{m}$ values for individual elements from the NIST XCOM database \cite{berger2010xcom}.  

For each soil, a composite $\mu_m$ was calculated from the elemental abundances (wt.\%) listed in Table~\ref{tab:soils}. The linear attenuation coefficient ($\mu$, \si{\per\cm}) was then calculated as:
\begin{equation}
\mu = \mu_m \cdot \rho_{\mathrm{total}} \label{eq:gamma_attenuation}
\end{equation}
where $\rho_{\mathrm{total}}$ is the total density of solids and liquids in the soil (dry solids + water).  

Applying a generic mass attenuation rate ($\mu_m$) to a soil of unknown composition requires the assumption that gamma attenuation at these energies is dominated by Compton scattering and pair production, both of which scale with electron density, and thus with total material density rather than specific composition \cite{tarim2013monte}. As we show later, $\mu_m$ varies little between soils, reinforcing the use of $\rho_{\mathrm{total}}$ as the primary control on gamma attenuation.

\begin{table*}
\centering
\resizebox{\linewidth}{!}{%
\begin{tblr}{
  column{3} = {c},
  column{4} = {c},
  column{5} = {c},
  column{6} = {c},
  column{7} = {c},
  cell{1}{3} = {c=4}{},
  cell{2}{2} = {c},
  cell{3}{2} = {c},
  cell{4}{2} = {c},
  cell{5}{2} = {c},
  cell{6}{2} = {c},
  cell{7}{2} = {c},
  cell{8}{2} = {c},
  cell{9}{2} = {c},
  cell{10}{2} = {c},
  cell{11}{2} = {c},
  cell{12}{2} = {c},
  cell{13}{2} = {c},
  cell{14}{2} = {c},
  cell{15}{2} = {c},
  cell{16}{2} = {c},
  cell{17}{2} = {c},
  cell{18}{2} = {c},
  hline{1,19} = {-}{0.08em},
  hline{2} = {3-7}{},
  hline{3,8} = {-}{},
}
                         &                                                        & Synthetic soils &           &            &           & Franz et al., (2013) soils (n=33) \\
Bulk properties          & Units                                                  & Dry porous      & Dry dense & Wet porous & Wet dense & Average (min - max)               \\
Dry bulk density         & g soil cm\textsuperscript{3} total                     & 1.1             & 1.5       & 1.1        & 1.5       & 1.313 (0.885 - 1.577)             \\
Volumetric water content & cm\textsuperscript{3 }H\textsubscript{2}O cm\textsuperscript{-3} total & 0   & 0     & 0.4        & 0.4       & 0.235 (0.051 - 0.481)             \\
Lattice water content    & g H\textsubscript{2}O g\textsuperscript{-1} soil       & 0.037           & 0.037     & 0.037      & 0.037     & 0.037 (0.003 - 0.064)             \\
Organic carbon content   & g C g\textsuperscript{-1} soil                         & 0.016           & 0.016     & 0.016      & 0.016     & 0.016 (0.002 - 0.081)             \\
Molar fractions          &                                                        &                 &           &            &           &                                   \\
\textsuperscript{16}O    & mol mol\textsuperscript{-1} total                      & 0.521           & 0.521     & 0.616      & 0.596     & 0.574 (0.523 - 0.634)             \\
\textsuperscript{28}Si   & mol mol\textsuperscript{-1} total                      & 0.327           & 0.327     & 0.242      & 0.260     & 0.282 (0.173 - 0.426)             \\
\textsuperscript{27}Al   & mol mol\textsuperscript{-1} total                      & 0.058           & 0.058     & 0.043      & 0.046     & 0.049 (0.001 - 0.091)             \\
\textsuperscript{56}Fe   & mol mol\textsuperscript{-1} total                      & 0.032           & 0.032     & 0.023      & 0.025     & 0.027 (0.004 - 0.072)             \\
\textsuperscript{39}K    & mol mol\textsuperscript{-1} total                      & 0.007           & 0.007     & 0.005      & 0.006     & 0.006 (0 - 0.021)                 \\
\textsuperscript{23}Na   & mol mol\textsuperscript{-1} total                      & 0.009           & 0.009     & 0.006      & 0.007     & 0.007 (0 - 0.021)                 \\
\textsuperscript{40}Ca   & mol mol\textsuperscript{-1} total                      & 0.012           & 0.012     & 0.009      & 0.009     & 0.010 (0 - 0.074)                 \\
\textsuperscript{24}Mg   & mol mol\textsuperscript{-1} total                      & 0.006           & 0.006     & 0.004      & 0.005     & 0.005 (0 - 0.014)                 \\
\textsuperscript{48}Ti   & mol mol\textsuperscript{-1} total                      & 0.004           & 0.004     & 0.003      & 0.003     & 0.003 (0.001 - 0.008)             \\
\textsuperscript{12}C    & mol mol\textsuperscript{-1} total                      & 0.019           & 0.019     & 0.014      & 0.015     & 0.015 (0.002 - 0.058)             \\
\textsuperscript{1}H     & mol mol\textsuperscript{-1} total                      & 0.006           & 0.006     & 0.033      & 0.027     & 0.021 (0.006 - 0.042)             
\end{tblr}
}
\caption{Bulk properties and molar abundances of selected isotopes used for MCNP simulations of soils. Four synthetic soils span a broad range of bulk density and water content, while lattice water and organic carbon are set to the mean of observations from Franz et al., 2013. The mean and range (minimum to maximum) for the observations from Franz et al., 2013 are shown in the rightmost column.}
\label{tab:soils}

\end{table*}

\subsection{Soil dataset and synthetic soils}
We used a dataset \citep{franz2013universal} that provides a complete set of soil properties for 33 U.S.\ soils, including dry bulk density ($\rho_{bd}$, in \si{\g} per \si{\cm\cubed} of soil), volumetric water content (VWC, $\theta$, in \si{\cm\cubed} of water per \si{\cm\cubed} of soil), lattice water content (LWC, in \si{\g} water per \si{\g} of soil), and organic carbon (in \si{\g} C per \si{\g} of soil). These soils span a wide range of textures, mineralogies, and organic matter contents. This dataset is valuable because accurate neutron transport modeling requires a full accounting of hydrogen sources. While bulk density and VWC are often measured, LWC and organic matter content are rarely reported together, yet both contribute hydrogen atoms that strongly affect neutron moderation \cite{wielopolski2005basic}.  

In addition to these natural soils, we evaluated four “synthetic soils” to isolate the effects of $\rho_{bd}$ and $\theta$. Modeled synthetic soils had constant LWC and organic matter (set to the mean of the dataset) but varied $\rho_{bd}$ and $\theta$ in increments of \qty{0.4}{\g\per\cm\cubed} and \qty{0.4}{\cm\cubed\per\cm\cubed}, respectively. This allowed for direct comparison of the attenuation effects of solid vs.\ water mass increases at equal density increments.

In Table~\ref{tab:soils} we present a summary of the soil properties of the modeled soils, as well as the elemental abundances of the soil mixture used as the MCNP input.

\subsection{Accounting for hydrogen sources}
Hydrogen content in soils was calculated from three components:
\begin{enumerate}
    \item Free water: reported as VWC ($\theta$, volume of water per volume of soil), converted to water mass per volume of soil using water density. This factor may change rapidly with time, depending on rainfall amounts and soil water retention characteristics, but evaluated soils primarily lie within the range of \numrange{0}{0.5}.
    \item Lattice water: hydroxyl groups bound in mineral lattices, expressed as LWC (g H$_2$O per g solid mineral). Lattice water ranges from 0-6\% in the evaluated soils.
    \item Organic matter hydrogen: estimated from organic carbon content, assuming cellulose composition (C$_6$H$_{10}$O$_5$) \cite{wielopolski2005basic}, which has a high H:C ratio and thus represents a maximum plausible hydrogen content from organic matter.
\end{enumerate}

This explicit accounting is important because hydrogen’s large scattering cross section makes it the dominant moderator of fast neutrons. Even small errors in hydrogen estimation can lead to significant neutron attenuation prediction errors.

\subsection{MCNP simulation setup}
We used MCNP 6.1 \cite{goorley2012initial} with ENDF/B-VIII.0 nuclear data libraries \cite{brown2018endf} to simulate neutron and gamma transport. The geometry was a homogeneous \qtyproduct{200x200x100}{\cm} soil block, with an isotropic \qty{14.1}{\MeV} neutron source located \qty{60}{\cm} above the surface. Each soil simulation used \qty{2e9} neutrons. We did not model detector housings or shielding, focusing solely on particle transport in soil. The chemical composition of a simulated material was designated by weight percents of individual isotopes in the MCNP input file, and the total density was specified separately. Soil is a mixture of different components: we added together the chemical compositions of the solid material, water content, and organic matter into a scaled weight percent array for each element. We simulated only 11 elements and only the major stable isotope for each (e.g., $^{16}$O), because trace elements (e.g., Cd or Zn) or rare stable isotopes (e.g., $^{18}$O) are typically less abundant by orders of magnitude in natural soils.

Gamma production from $^{12}$C(n,n’$\gamma$) was tallied using an F4 mesh tally with FM multipliers for the reaction cross section, in \qty{1}{\cm} vertical bins down to \qty{60}{\cm} depth. From the results, we calculated the INS reaction rate given in \unit{gammas \per \cm\cubed\per(source\,neutron)}. The FM tag enables transformation of the neutron flux from the F4 tally into the reaction rate (RR) of INS reaction \cite{doron2007simulation}. 

\begin{equation}
RR_i = N_i \int_0^\infty \!\!\sigma_i(E) \,\phi(E) \,dE
\end{equation}

where $\phi$ is the neutron flux, $\sigma_i$ is the energy-dependent cross-section for a neutron collision with some element $i$. N$_i$ the number density of the subject atom, calculated as

\begin{equation}
N_i = \frac{w_i \,\rho_{total} \,N_A}{M_i}
\end{equation}

where $w_i$ is the weight fraction the element $i$ in the modeled soil material, $\rho_{total}$ is the total density (g cm$^{-3}$), $N_A$ is Avogadro's number (0.6022 atoms cm$^{2}$ mol$^{-1}$ barn$^{-1}$) so that the number density N$_i$ has units of atoms barn$^{-1}$ cm$^{-1}$.

We calculated the reaction rate for INS reactions with C that produce \qty{4.43}{\MeV} gamma rays in the superimposed mesh tally structure grouped into \qty{1}{\cm\cubed} voxel cells. The output file \texttt{meshtal} reports the reaction rate for cells of different positions $X$, $Y$, $Z$ as well as energy bins that correspond to the energy of the incident neutron. To observe the attenuation of the INS reaction rate through the soil column, we summed reaction rates in each voxel across $X$ and $Y$ to yield the change in reaction rate with depth $Z$. For comparison of soils with different amounts of the INS subject element (i.e., carbon-rich or carbon-poor soils), we normalized reaction rates calculated at depth to that of the surface at \qty{0}{\cm}. This allowed for comparison of fractional attenuation of INS reaction rates with depth. 

Application of the INS-API method involves the assumption that the initial neutron travel direction (as inferred from the alpha detector) is not modified prior to generating a gamma ray via the INS reaction. In other words, all detected gamma rays are generated via scattering reactions by neutrons that have not had any collisions that dramatically alter the direction of travel. It is likely that \qty{14.1}{\MeV} neutrons may undergo one or more collisions which lower the neutron energy to a small enough extent that it may later generate a gamma ray (i.e., if it stays above the minimum energy for incident neutrons to undergo an INS reaction, \qty{5}{\MeV} for the case of carbon). For a rough approximation of the rate of INS reactions induced by nominally ``uncollided'' neutrons, a narrow energy bin of \qtyrange{14.0}{14.1} MeV was set. Since the average decrement of energy per collision of \qty{14.1}{\MeV} neutrons in soil is much higher than \qty{0.1}{\MeV} \cite{spielberg1958dose}, this bin likely encompassed all neutrons which have not collided and reduced in energy from \qty{14.1}{\MeV}. We summed reaction rates for all incident energies to represent the total reaction rate. We then compared ``uncollided'' (\qtyrange{14.0}{14.1}{\MeV}) with the total reaction rate vs. depth to determine the approximate contribution of each group to the INS reaction rate trends observed.

\subsection{Fitting attenuation curves}

To generate a first-order prediction of the attenuation of \qty{4.43}{\MeV} gamma reaction rates with depth in soil, we fit a function to describe the normalized reaction rate (relative to the surface) at different depths for each modeled soil. The exponential function is useful for describing the attenuation of a mono-energetic particle (e.g., attenuation of gamma rays):

\begin{equation}
RR_{norm}(z) = a \cdot \exp\left(\frac{-z}{b}\right)
\label{eq:singleexp}
\end{equation}

However, the INS reaction for C that generates \qty{4.43} {\MeV} gamma rays can be initiated by neutrons of varying energies. We found that a double-exponential function is more appropriate to represent the attenuation of the C INS reaction rate:

\begin{equation}
RR_{norm}(z) = a \cdot \exp\left(\frac{-z}{b}\right)+  (1-a) \cdot \exp\left(\frac{-z}{c}\right)
\label{eq:doubleexp}
\end{equation}
where $a$ is the fraction of the fast-attenuating component, and $b$ and $c$ are attenuation lengths for the fast and slow components, respectively.

First, we fit the double-exponential function to the reaction rate data vs.\ depth to obtain attenuation parameters ($a$,$b$, and $c$) for all modeled soils. We then regressed $a$, $b$, and $c$ against the bulk density ($\rho_{bd}$) and water content ($\theta$):
\begin{equation}
i = d_i \, \rho_{bd} + e_i \, \theta + f_i
\label{eq:regression}
\end{equation}
where $i \in \{a, b, c\}$ and $d_i$, $e_i$, and $f_i$ are fitted constants.  

After generating predictive estimates of neutron attenuation by fitting to bulk density ($\rho_{bd}$) and soil water content ($\theta$), we explored whether other sources of water (e.g., lattice water or organic water) are sufficiently abundant to require additional consideration. For this, we also tested effective $\rho_{bd}^*$ and $\theta^*$ values including lattice and organic water in Equation \ref{eq:regression}:
\begin{equation}
 \rho _{bd}^{*}=\rho_{bd}\cdot(1-LW-OW)
 \label{eq:effective_bd}
\end{equation}
\begin{equation}
  \theta^{*} = \theta + \rho _{bd} \cdot LW + \rho_{bd} \cdot M_{OM} \cdot OW 
   \label{eq:effective_theta}
\end{equation}
where $LW$ is lattice water fraction, $M_{OM}$ is organic matter mass fraction, and $OW$ is organic water fraction (from cellulose stoichiometry).

\subsection{Experimental validation}
We validated the model using the INS–API system described in \cite{ayllon2021all, unzueta2019associated}, with a DT108API neutron generator (Adelphi Technologies Inc.), LaBr$_3$ (3" Saint-Gobain, B390S) and NaI (5" Alpha Spectra, Inc. 20I20/5(9823)BN) gamma detectors, and a YAP scintillator (Crytor Inc.) and a position-sensitive photo multiplier tube (Hamamatsu H13700-03).  

Two experiments were conducted to validate the neutron attenuation model:
First, the effects of moisture variation were evaluated by adjusted volumetric water content ($\theta$) to 2.6, 11.1, 22.8, 33.1\%, given a fixed soil depth of \qty{10}{\cm}. Second, attenuation with depth was evaluated for dry soil ($\theta$ = 2.6\%) and a soil depth range of \qtyrange{3}{18}{\cm}.

\begin{figure}[htb]
\centering
\includegraphics[width=0.8\linewidth]{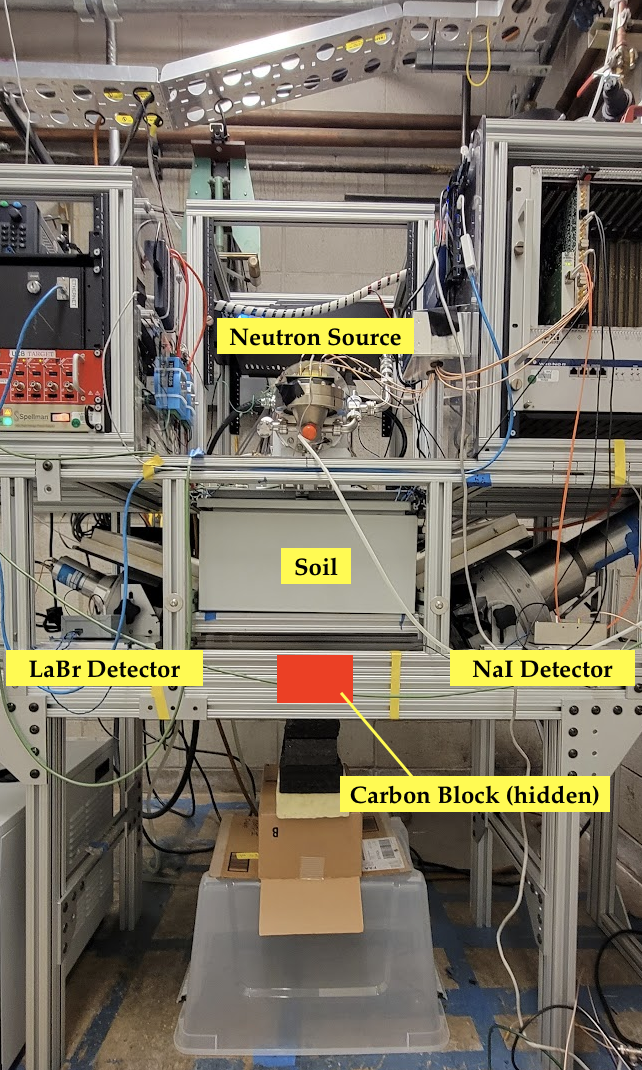}
\caption{Experimental setup for neutron attenuation measurements. A steel soil container is placed between the neutron source and a graphite block target. Soil depth and moisture were varied to assess their effect on neutron attenuation.}
\label{fig:setup}
\end{figure}

The setup is shown in Figure~\ref{fig:setup}. To avoid complications from additional gamma attenuation and to focus on signals of C in soil (our main interest), we placed a container with soil of varying thickness and moisture level above a graphite block and looked for API signals from the graphite block only. The soil container changed the neutron flux hitting the graphite block, but since we did not have any soil between the graphite block and the gamma detectors, the gamma signal was not attenuated.
Using a steel container minimized gamma interference at the carbon energy of \qty{4.43}{\MeV} compared to using a plastic container. Background from the empty container was subtracted from each run to minimize the influence of the container itself (this method is not perfect, since we acknowledge that the neutron flux on the container will vary with the soil used and between a measurement with soil and without soil).

The soil used was sandy loam (65.8\% sand, 18.0\% silt, 16.2\% clay). Prior to its use in these experiments, the soil was sifted to $<$\qty{2}{\mm} for large pebbles and roots, making it largely uniform in composition and geometry. The initial moisture content of the soil was measured by weighing six \qty{8}{\g} samples of soil drawn from the sifted source soil, oven drying the samples at \qty{105}{\celsius} for 16 hours, then weighing them again. We used the average of these values for the initial moisture content of the bulk soil container (\qty{2.3}\% gravimetric water content, \qty{2.6}\% volumetric water content).

In order to measure the attenuating effect of moisture in soil on API measurements, we placed a box filled with soil moistened to four incremental water contents in between the \qty{14}{\MeV} neutron source and a target carbon block. We chose a water-tight, \qtyproduct{40 x 40 x 20}{\cm} box made from \qty{1.25}{\mm} thick steel to hold the soil. The iron-rich steel composition minimizes \qty{4.43}{\MeV} scattering interference with the carbon target compared to a plastic container; however, to account for the interference from the box, we first took a measurement with only the empty steel box above the carbon block, which we later used as a baseline for normalization of the gamma counts.

We then filled the box \qty{10}{\cm} high with sifted soil and added water using a water misting apparatus which gradually and uniformly moistened the soil. We used a scale to measure the total amount of water added. We tested incremental volumetric water contents of 11.1\%, 22.8\%, and 33.1\%. At each increment, we sealed the container and waited one hour for the water to percolate uniformly into the soil. As soon as the moisture content became uniform, we placed and centered the soil container under the neutron source with a distance of \qty{27}{cm} between the neutron production target and the top of the container. To isolate the effect of attenuating material depth on neutron attenuation, we took additional measurements at varying depths of sifted dry (2.6\% VWC) soil. We measured depths of \qty{3}{\cm}, \qty{5}{\cm}, \qty{15}{\cm}, and \qty{18}{\cm} and used the same spatial set-up as for the moist soil measurements. 

For each measurement, we took one two-hour measurement of just the box containing the soil, and one two-hour measurement of a \qtyproduct{20 x 13 x 6}{\cm} graphite block centered \qty{13}{cm} under the bottom of the soil container (Figure~\ref{fig:setup}). The carbon block was placed on low-density foam to limit errors in measurement caused by the carbon in the foam. The first two-hour run provided background scattering from the attenuating material itself that we later subtracted from the measured \qty{4.43}{\MeV} carbon gamma energy peak. 

\begin{figure}
\includegraphics[width=0.9\linewidth]{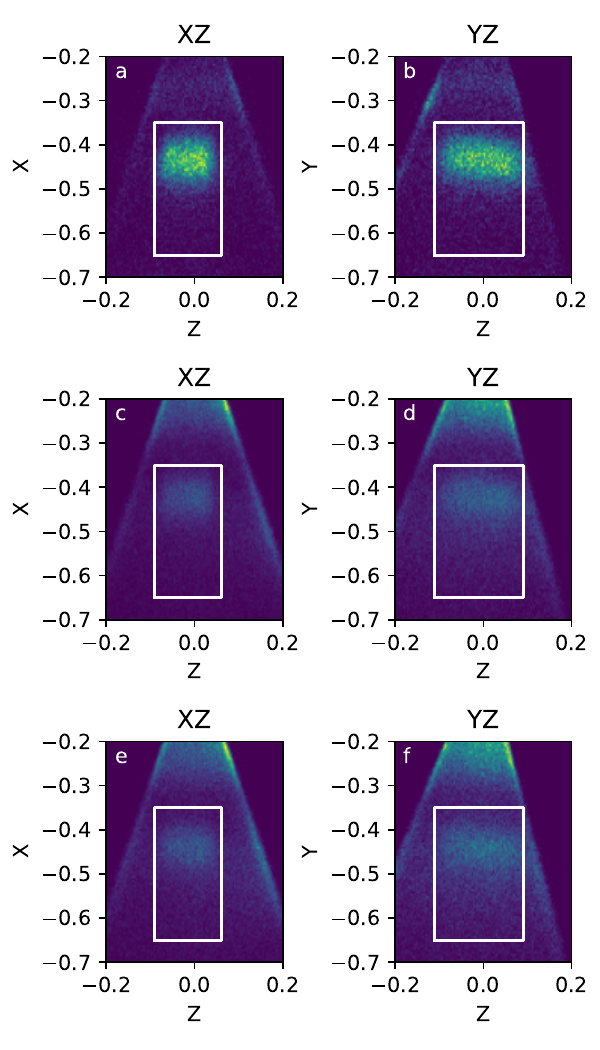}
\caption{Voxelized $XZ$ and $YZ$ cross-sections of the reconstructed carbon block signal. (a,b) Steel box only; (c,d) 10 cm soil at 40\% VWC; (e,f) \qty{18}{\cm} dry soil. Voxel boundaries capture the carbon block signal without crossing the API cone edge.}
\label{fig:voxel_image}
\end{figure}

After the measurements were completed, we estimated the neutron flux going into the carbon block by integrating the \qty{4.43}{\MeV} carbon gamma energy peak, including the first escape peak and the Compton edge, over the volume of the block (Figure ~\ref{fig:voxel_image}). The voxel shape was determined by maximizing the area in $XZ$ and $YZ$ that did not cross over the edge of the API cone. We extended the volume of the voxel in the negative $Z$ direction to capture scattering from the carbon block. We normalized this value to the number of alpha particles detected in the run. Alpha particles are taken as a 1:1 proxy for generated neutrons. We did the same for the background measurement and subtracted it from the run with the carbon block to remove background scattering from measurements of carbon counts. This produced a value of normalized neutron flux for each moisture increment. Percent neutron attenuation was calculated as the normalized neutron flux for each moisture content divided by the normalized neutron flux for the empty steel box measurement.

\section{Results}

\subsection{Simulation results}
\begin{figure}[htp]
\centering
\includegraphics[width=0.8\linewidth]{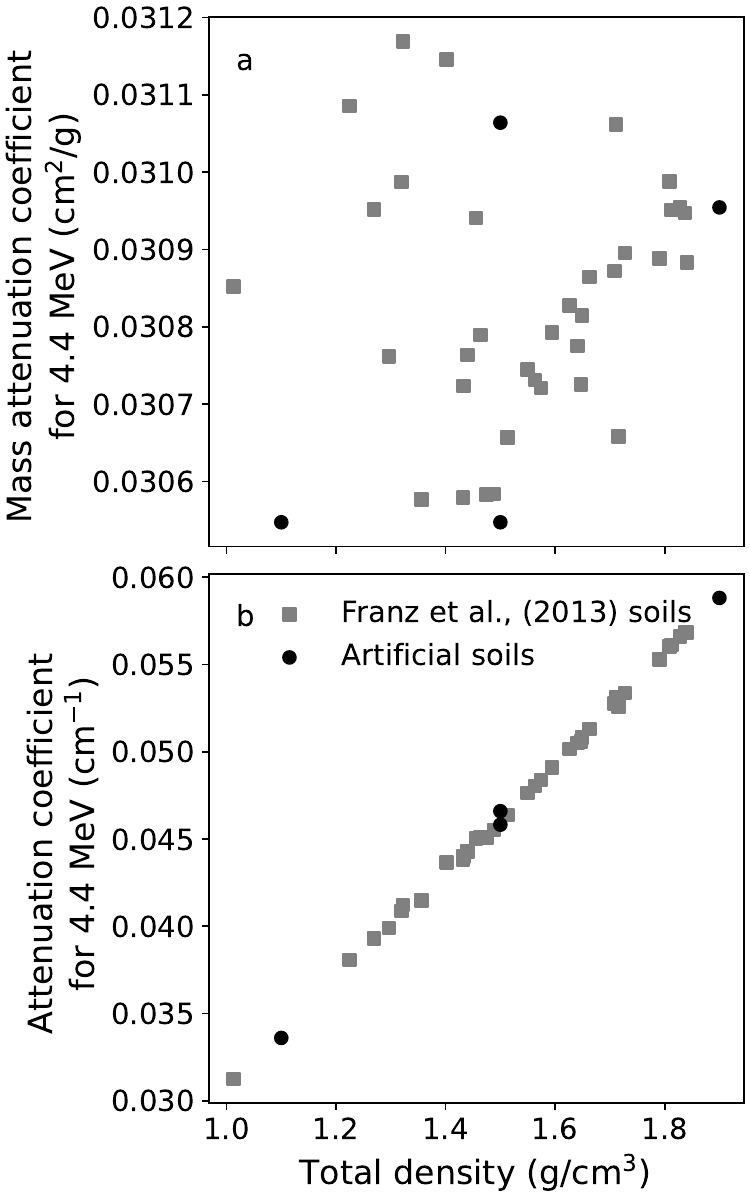}
\caption{Mass attenuation coefficients for a \qty{4.43}{\MeV} gamma ray plotted vs. total density of solids and liquids in soil (a). Multiplication by total density yields the linear attenuation coefficient (b).}
\label{fig:gamma}
\end{figure}

The attenuation of high-energy ($>\sim$\qty{1}{\MeV}) gamma rays through soil shows minimal dependence on soil composition. For all soils tested, the mass attenuation coefficient $\mu_{m}$ for \qty{4.43}{\MeV} gamma rays varied little (Figure~\ref{fig:gamma}a). This reflects the fact that at these energies, Compton scattering dominates, and cross sections are relatively insensitive to atomic number for the light and medium elements common in soil \cite{tarim2013monte}.  

\begin{table*}
\centering
\begin{tblr}{
  cells = {c},
  cell{1}{1} = {r=3}{},
  cell{1}{2} = {c=6}{},
  cell{2}{2} = {c=3}{},
  cell{2}{5} = {c=3}{},
  hline{1,4,7} = {-}{},
  hline{2-3} = {2-7}{},
}
{Reaction rate\\attenuation constants  } & Regression constants            &         &        &                                            &         &        \\
                                         & Simple $\rho_{bd}$ and $\theta$ &         &        & Effective $\rho_{bd}^{*}$ and $\theta^{*}$ &         &        \\
                                         & $d$                             & $e$     & $f$    & $d$                                        & $e$     & $f$    \\
$a$                                      & -0.103                          & -0.469  & 0.081  & -0.107                                     & -0.446  & 0.101  \\
$b$ (cm)                                      & -0.038                          &  3.248  & 2.631  & -0.022                                     & 3.067   & 2.479  \\
$c$ (cm)                                     & -6.835                          & -15.343 & 27.559 & -6.976                                     & -14.813 & 28.063 \\
\end{tblr}

\caption{Results from regressing bulk density ($\rho_{bd}$) and water content ($\theta$) for 33 modeled soils \cite{franz2013universal} against their regression constants for the double exponential decay in Equation \ref{eq:doubleexp}. $\rho_{bd}$ and $\theta$ serve as two independent variables in the regression to predict decay constants $a$, $b$, and $c$, according to Equation \ref{eq:regression}. Additionally results for the effective values ($\rho_{bd}^{*}$ and $\theta^{*}$, as in Equation \ref{eq:effective_bd} and \ref{eq:effective_theta}) are shown.}
\label{tab:regressionresults}

\end{table*}

Because the linear attenuation coefficient $\mu$ is simply $\mu_{m}$ multiplied by the total density of material, $\mu$ is strongly correlated with $\rho_{\mathrm{total}}$ (Figure~\ref{fig:gamma}b). This confirms that gamma attenuation can be reliably estimated from density measurements alone, without detailed knowledge of composition, a technique that is also used, for example, in oil exploration \cite{Alger1963-bo,Borsaru1985-mj}.

The measured rate of C from INS reactions as a function of depth is more complex than pure gamma attenuation due to the soil between the origin of the gamma and the detector because \qty{4.43}{\MeV} gammas can be generated by neutrons over a range of incident energies (\qtyrange{4.8}{20}{\MeV}). Neutrons may undergo several collisions and lose energy before producing a gamma ray.  

\begin{figure*}
\centering
\includegraphics[width=0.7\linewidth]{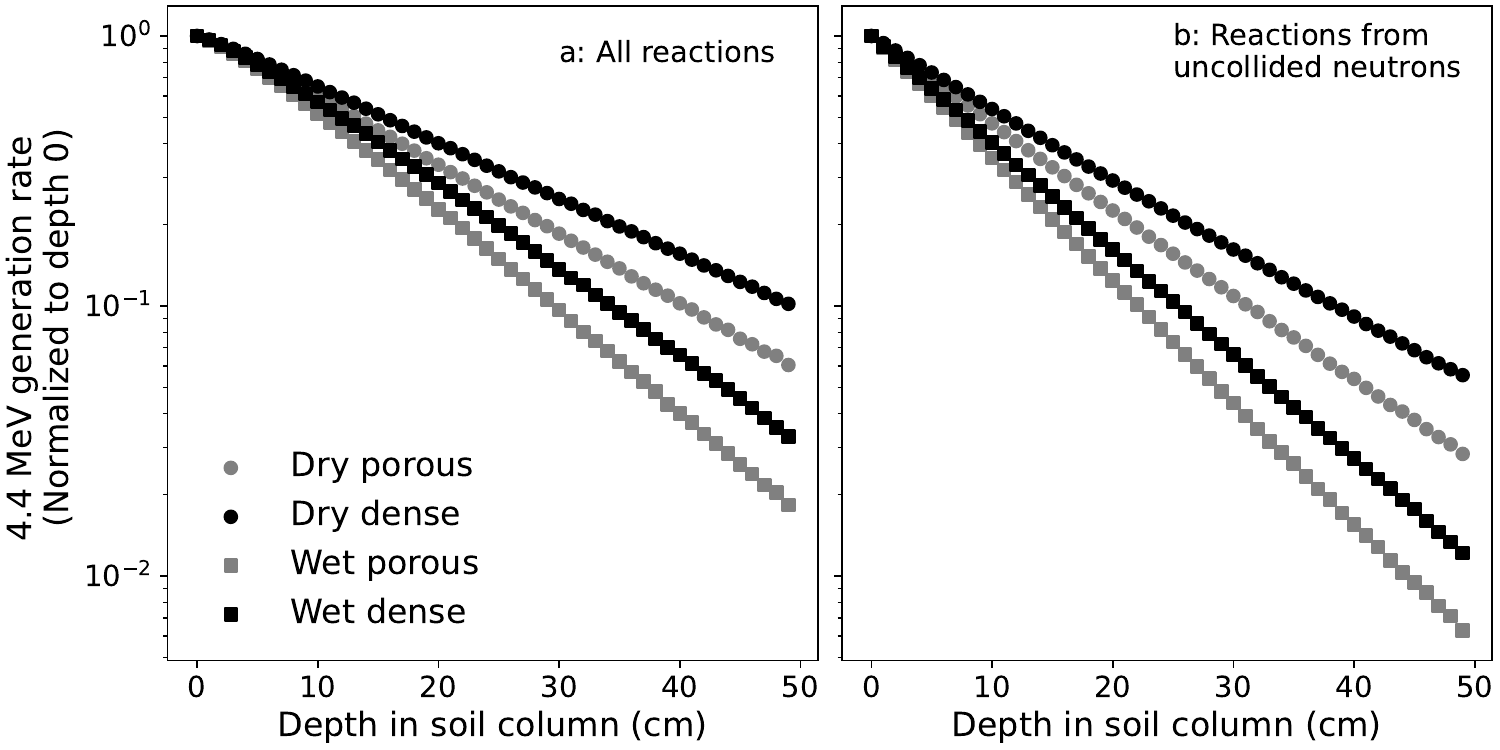}
\caption{Normalized \qty{4.43}{\MeV} gamma production vs. depth for four synthetic soils (a) and for uncollided \qty{14.1}{\MeV} neutrons only (b).}
\label{fig:rr}
\end{figure*}

Figure~\ref{fig:rr}a shows normalized reaction rates for four synthetic soils designed to isolate the effects of bulk density and water content. Increasing $\rho_{bd}$ by \qty{0.4}{\g\per\cm\cubed} (porous → dense) reduced the reaction rate moderately, but increasing $\theta$ by the same mass-equivalent density (\qty{0.4}{\g\per\cm\cubed}) had a much stronger effect. This illustrates hydrogen’s efficiency as a neutron moderator.  

For uncollided \qty{14.1}{\MeV} neutrons (Figure~\ref{fig:rr}b), attenuation follows a near-exponential decay, as expected for monoenergetic particles. The steeper slope in wetter soils reflects the rapid moderation of fast neutrons by hydrogen in water.

\begin{figure}
\centering
\includegraphics[width=0.8\linewidth]{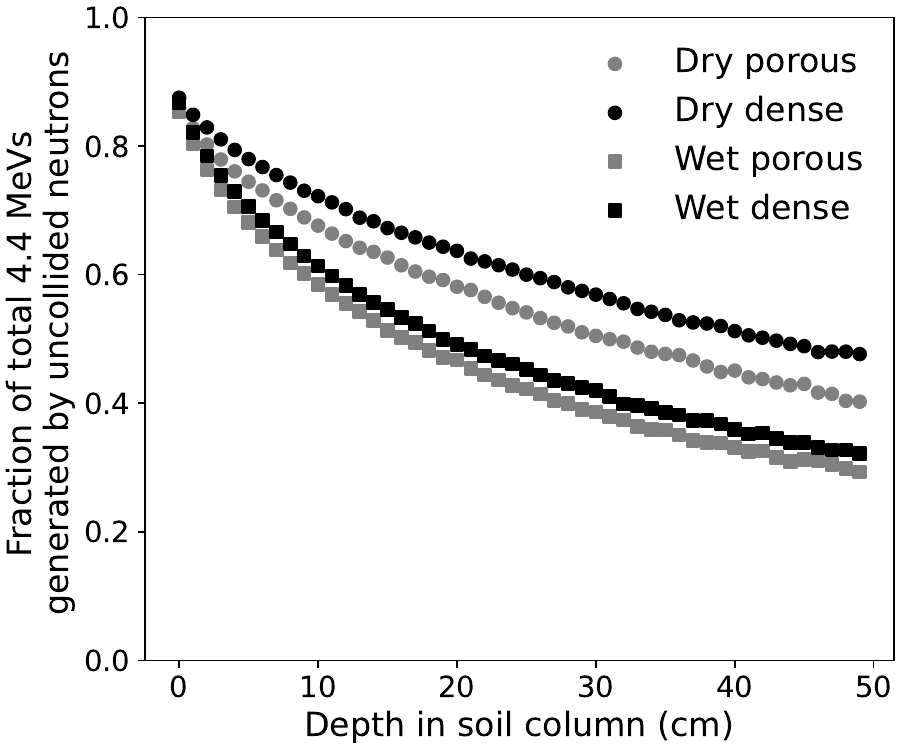}
\caption{Fraction of \qty{4.43}{\MeV} gamma rays from uncollided (\qtyrange{14.0}{14.1}{\MeV}) neutrons vs. depth for four synthetic soils.}
\label{fig:uncollided}
\end{figure}

The fraction of \qty{4.43}{\MeV} gamma rays produced by uncollided neutrons decreases with depth (Figure~\ref{fig:uncollided}). In wet soils, this drop is rapid, indicating that most gammas at depth are generated by neutrons that have already scattered and lost energy.  

This has implications for INS–API: while the alpha detector gives the initial neutron trajectory, multiple scattering events can alter that trajectory before gamma production, introducing spatial uncertainty in voxel reconstructions. Quantifying this effect would require detailed particle tracking (e.g., using MCNP's \texttt{ptrac} feature), which is beyond the scope of this paper.

\subsection{Regression model performance}
\begin{figure*}
\centering
\includegraphics[width=0.8\linewidth]{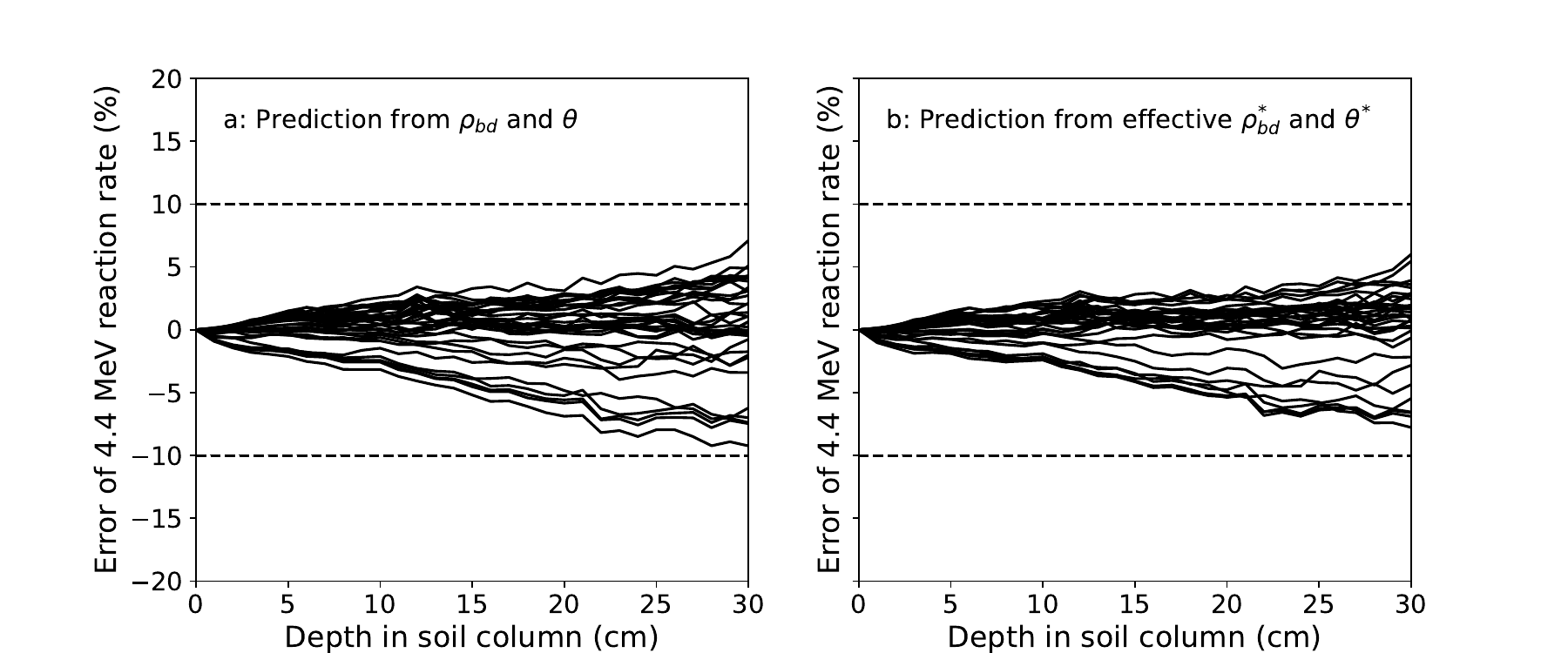}
\caption{Percent error between predicted and simulated \qty{4.43}{\MeV} gamma attenuation for 33 soils: (a) using $\rho_{bd}$ and $\theta$; (b) using effective $\rho_{bd}^*$ and $\theta^*$.}
\label{fig:8}
\end{figure*}

We fitted the double-exponential model (Equation~\ref{eq:doubleexp}) to simulated attenuation curves for 33 real soils \cite{franz2013universal}. Regression of $a$, $b$, and $c$ against $\rho_{bd}$ and $\theta$ (Equation~\ref{eq:regression}) predicted attenuation curves with percent errors generally $<\pm$10\% at \qty{30}{\cm} depth (Figure~\ref{fig:8}a) and smaller errors at shallower depth.  The result of the regression model is summarized in Table~\ref{tab:regressionresults}.

Including lattice and organic water in effective $\rho_{bd}^*$ and $\theta^*$ improved predictions slightly (Figure~\ref{fig:8}b), especially for clay-rich or organic-rich soils.  

Larger errors occurred in soils with extreme combinations of low $\rho_{bd}$ and high $\theta$, suggesting secondary influences from other elements (e.g., Fe) on neutron attenuation.

\subsection{Experimental results}
\begin{figure}
\centering
\includegraphics[width=0.9\linewidth]{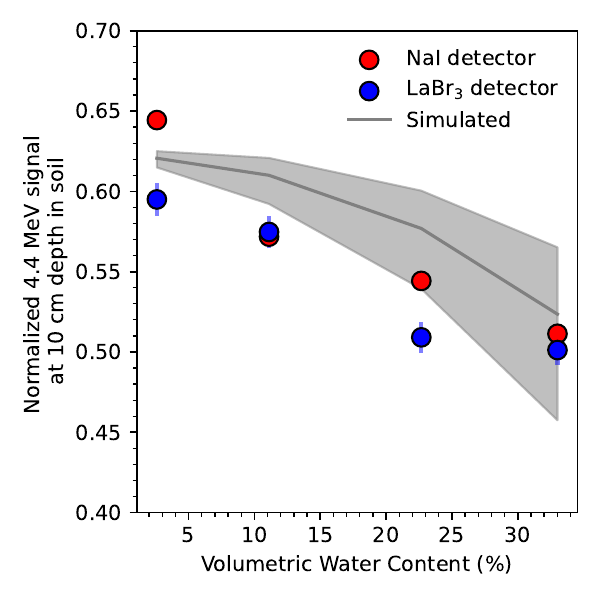}
\caption{Gray lines: model predictions for neutron attenuation at \qty{10}{\cm} depth for different VWC. Circles: experimental results from NaI (red) and LaBr$_3$ (blue) detectors. Error bars are around 1\% and include counting statistic only.}
\label{fig:results_moisture}
\end{figure}

In the fixed-depth experiment (\qty{10}{\cm}), normalized neutron flux into the graphite target decreased with increasing VWC (Figure~\ref{fig:results_moisture}). Model predictions matched measured attenuation trends for both NaI and LaBr$_3$ detectors.  

This agreement supports the model’s ability to predict neutron attenuation from simple field-measurable parameters.

\begin{figure}[htp]
\centering
\includegraphics[width=0.9\linewidth]{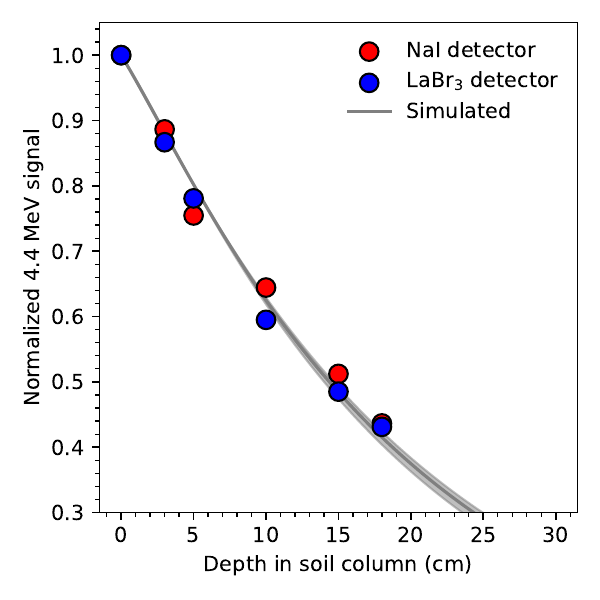}
\caption{Gray line: model prediction for dry soil (2.6\% VWC). Circles: experimental results from NaI (red) and LaBr$_3$ (blue) detectors. Error bars are around 1\% and include counting statistic only.}
\label{fig:result_depth}
\end{figure}

In the dry-soil depth experiment ($\theta$ = 2.6\%), neutron flux decreased with depth in agreement with model predictions (Figure~\ref{fig:result_depth}). Agreement was within $\sim$10\% at \qty{30}{\cm} depth, consistent with the simulation-based error estimates.

\section{Discussion}

\subsection{Dominance of water content in neutron attenuation}
Our simulations and experiments show that volumetric water content ($\theta$) is the dominant factor controlling neutron attenuation in soils. This is consistent with the high scattering cross section of hydrogen \cite{wielopolski2005basic}, which efficiently moderates fast neutrons. Even modest increases in $\theta$ substantially reduce the population of fast neutrons capable of inducing the $^{12}$C(n,n’$\gamma$) reaction at \qty{4.43}{\MeV}.  

The synthetic soil simulations demonstrate that an increase in water content equivalent in mass density compared to an increase in just solid material has a far greater attenuating effect. This is because neutron moderation is driven primarily by elastic scattering with light nuclei, and hydrogen is the lightest and most efficient moderator.

\subsection{Role of solid soil material}
Dry bulk density ($\rho_{bd}$) also influences neutron attenuation by increasing the number of nuclei per unit volume, thereby increasing the probability of neutron interactions. However, without hydrogen’s moderating effect, the impact is smaller than that of water content. This aligns with earlier neutron transport studies in soils \cite{wielopolski2008nondestructive}.

\subsection{Gamma-ray attenuation vs. neutron attenuation}
Gamma-ray attenuation at \qty{4.43}{\MeV} depends primarily on total density and is relatively insensitive to soil composition within the range studied \cite{tarim2013monte}. This contrasts with neutron attenuation, which is highly sensitive to hydrogen content.  

As a result, gamma attenuation corrections can be applied using tabulated mass attenuation coefficients and total density measurements, whereas neutron attenuation requires site-specific modeling or empirical correction.

\subsection{Model practicality and limitations}
The regression model developed here requires only $\rho_{bd}$ and $\theta$ — quantities that can be measured in the field using standard soil sampling or in situ sensors \cite{Topp1980-ll,Evett2008-fg}. Including lattice and organic water in effective $\rho_{bd}^*$ and $\theta^*$ slightly improves precision, particularly in clay-rich or organic-rich soils \cite{wielopolski2005basic}.  Both bulk density of soil solids and water content can however also be modeled directly from the measured data enabling a self-consistent framework that allows non-destructive sampling. An additional neutron detector or running the neutron generator in pulsed mode for a short time can allow the direct measurement of H concentration in the soil leading to better moisture estimates \cite{Yakubova2025-yu}. Together with the model presented here, the INS-API data can then be used to provide accurate elemental estimates without relying on additional measurements.

Importantly, the current model of signal attenuation in soils assumes homogeneous soil properties with depth. In layered soils or those with significant contrasts in moisture or texture, attenuation may deviate from predictions. Our validation experiments were conducted under controlled laboratory conditions using a single soil type; a larger field validation is needed for diverse soils and environmental conditions.

\section{Conclusions}

By enabling accurate correction for neutron attenuation, our model improves the quantitative accuracy of voxelated carbon measurements from INS–API systems \cite{unzueta2019associated,ayllon2021all}. This is particularly important for carbon accounting in agricultural and rangeland soils, where management practices can alter both $\rho_{bd}$ and $\theta$.  
The approach also has potential applications in other neutron-based analyses, such as nitrogen detection for explosive detection in soil \cite{Vourvopoulos2001-f9a,Faust2011-gj}. 

We developed and validated an empirical model that predicts neutron attenuation in soils from only dry bulk density and volumetric water content. Monte Carlo simulations across 33 real and 4 synthetic soils showed that water content has a much stronger effect on neutron attenuation than bulk density. Attenuation profiles were well described by a double-exponential function, and regression models predicted attenuation within $\sim$10\% at \qty{30}{\cm} depth for most soils.  

Experimental validation using an INS–API system confirmed the model’s accuracy under controlled conditions. Future work will focus on validating the complete framework including bulk density and water estimation from INS-API data. This model enables practical, field-deployable correction of INS–API data for neutron attenuation, laying the groundwork for a self-consistent framework that improves the reliability of spatially resolved soil carbon measurements. 

\section*{Acknowledgements}
We thank Takeshi Katayanagi for mechanical support. 

\section*{Funding}
The information, data, or work presented herein was funded by the Advanced Research Projects Agency-Energy (ARPA-E), U.S. Department of Energy, and the U.S. Department of Energy (DOE) through Lawrence Berkeley National Laboratory's Laboratory-Directed Research and Development (LDRD) as part of its Carbon Negative Initiative, under Contract No. DEAC02-05CH11231. This work was supported in part by the U.S. Department of Energy, Office of Science, Office of Workforce Development for Teachers and Scientists (WDTS) under the Science Undergraduate Laboratory Internship (SULI) program.

\section*{CRediT authorship contribution statement}
\textbf{William Larsen}:  Conceptualization, Methodology, Software, Validation, Formal analysis, Investigation, Writing - Original Draft, Writing - Review \& Editing, Visualization
\textbf{Valerie Smykalov}:   Conceptualization, Methodology, Software, Validation, Formal analysis, Investigation, Writing - Original Draft, Visualization
\textbf{Cristina Castanha}: Resources, Methodology
\textbf{Eoin Brodie}:  Conceptualization, Supervision, Funding acquisition
\textbf{Mauricio Ayllon Unzueta}: Conceptualization, Methodology, Software, Validation, Formal analysis
\textbf{Bernhard Ludewigt}: Conceptualization, Methodology, Investigation, Supervision
\textbf{Arun Persaud}: Conceptualization, Methodology, Software, Validation, Formal analysis, Investigation, Resources, Writing - Review \& Editing, Supervision, Project administration, Funding acquisition

\section*{Data availability}
Raw data, analysis scripts including some additional plots, and MCNP input files are available at Zenodo \cite{larsen_2026_18133672}.

\section*{Declaration of competing interest}
The authors declare that they have no known competing financial interests or personal relationships that could have influenced the work reported in this paper. 
However, it is important to disclose that Author William Larsen is currently employed by Terradot, a company focused on carbon sequestration. This research was conducted during his tenure at Lawrence Berkeley National Lab and does not involve any products or services of Terradot. The authors affirm that this affiliation does not represent a conflict of interest regarding the findings presented in this manuscript.

\section*{Declaration of generative AI and AI-assisted technologies in the writing process}
During the preparation of this work the author(s) used openAI:chatgpt5 in order to improve the readability of the article. After using this tool/service, the authors reviewed and edited the content as needed and take full responsibility for the content of the published article.

\bibliographystyle{elsarticle-num}
\bibliography{paper}

\end{document}